\documentclass[aps, prb , reprint,superscriptaddress]{revtex4-2}
\usepackage{verbatim}
\usepackage{bbold}
\usepackage{amsmath}
\usepackage[english]{babel}
\usepackage{graphicx}
\usepackage{mathtools}
\usepackage{hyperref} 
\usepackage{textcomp}% for the \textmu
\usepackage{subcaption}
\usepackage{graphicx}
\usepackage{lipsum}
\usepackage[normalem]{ulem}
\hypersetup{colorlinks=true, linkcolor=blue, citecolor=blue, urlcolor=blue}

\begin{document}
\title{Chiral Quantum Optics with Scalable Quantum Dot Dimers}

\author{L. Hallacy}
\email[]{lmhallacy1@sheffield.ac.uk}
\affiliation{School of Mathematical and Physical Sciences, University of Sheffield, Sheffield, S3 7RH, UK}

\author{D. Hallett}
\affiliation{School of Mathematical and Physical Sciences, University of Sheffield, Sheffield, S3 7RH, UK}

\author{A. Fenzl}
\affiliation{School of Mathematical and Physical Sciences, University of Sheffield, Sheffield, S3 7RH, UK}

\author{N.J. Martin}
\affiliation{School of Mathematical and Physical Sciences, University of Sheffield, Sheffield, S3 7RH, UK}

\author{R. Dost}
\affiliation{School of Mathematical and Physical Sciences, University of Sheffield, Sheffield, S3 7RH, UK}

\author{A.Verma}
\affiliation{School of Electrical and Electronic Engineering, University of Sheffield, Sheffield, S1 3JD, UK}

\author{J. Fletcher}
\affiliation{Ion Beam Centre, University of Surrey, Guildford, UK, GU2 7XH}

\author{I. Farrer}
\affiliation{School of Electrical and Electronic Engineering, University of Sheffield, Sheffield, S1 3JD, UK}

\author{L. Antwis}
\affiliation{Ion Beam Centre, University of Surrey, Guildford, UK, GU2 7XH}

\author{M.S. Skolnick}
\affiliation{School of Mathematical and Physical Sciences, University of Sheffield, Sheffield, S3 7RH, UK}

\author{L.R. Wilson}
\affiliation{School of Mathematical and Physical Sciences, University of Sheffield, Sheffield, S3 7RH, UK}

\date{\today}
\begin{abstract}
We present a scalable method for electrically tuning multiple spatially separated quantum dots embedded in photonic crystal waveguides. Ion implantation into the top p-doped layer of a \textit{p--i--n} diode creates high-resistivity tracks, providing electrical isolation between adjacent regions. Unlike physical etching, this method preserves the guided-mode profile of the photonic crystal without introducing significant scattering, limiting refractive index perturbations to below $10^{-3}$ with  $ 0.01\% $ additional loss. In contrast, physical etching can reduce single-band transmission by more than $ 30\% $ for a etch width of $100$ nm. We demonstrate the applicability of our approach using quantum dots embedded in a glideplane photonic crystal waveguide, controlling the detuning between different spin-state combinations of two highly chiral quantum dots coupled to the same mode. Second-order photon correlation measurements provide a sensitive probe of the chirality-dependent photon statistics, which are in good agreement with a waveguide–QED master equation model. Our results mark an important step towards scalable, multi-emitter architectures for chiral quantum networks.
\end{abstract}

\maketitle

\section{Introduction}

Chiral quantum networks offer a scalable route to distributed quantum information processing, enabling deterministic transfer of quantum states and photon-mediated entanglement between distant qubits \cite{mahmoodian2016quantum}. Such architectures underpin applications ranging from programmable quantum processors \cite{Madsen2022Borealis,Yang2025Programmable} and quantum simulators \cite{Slussarenko} to fault-tolerant quantum repeaters \cite{Niu2023}, where efficient and directional spin-photon interfaces form the key building block. Realizing these networks requires platforms that combine near-unity coupling efficiency with strong spin-selectivity while supporting scalable integration and fast tunability of multiple quantum emitters. Self-assembled QDs are highly controllable quasi-ideal quantum emitters in a solid-state platform \cite{Tomm2021, Somaschi2016, Meng2024}, making them ideal for building these systems. 
\\
Photonic crystal (PhC) waveguides provide several favorable properties for this task. Their periodic dielectric structure supports slow-light modes that enhance light-matter interaction, leading to simultaneously large $\beta$-factors, strong Purcell enhancement, and high chiral contrast \cite{Arcari2014NearUnity,Lodahl2015rmp,Sollner2015Deterministic,QuanLoncar2011}. A glide-plane waveguide (GPWG) is a particularly promising PhC structure \cite{germanis2025waveguideexcitationspinpumping}.
\\
In GPWGs, the broken symmetry ensures that circularly polarized emission from a QD couples robustly into a forward or backward propagating mode with low positional dependence on unidirectional emission. Recent comparative studies show that GPWGs outperform conventional W1 line-defect waveguides for achieving chiral coupling: while W1 structures typically yield only a small fraction of QDs with $>80\%$ chiral contrast, GPWGs exhibit the highest proportion of high-contrast QDs, with both simulations and experiments demonstrating robust regions where $\beta \gtrsim 80\%$ and chiral contrast exceeds $90\%$ \cite{Martin2023Topological,Hauff2022Chiral}. The low positional dependence of GPWGs is especially attractive for embedded QDs, which inhibit inhomogeneous emission energies and random spatial positions \cite{SKgrowth,Zhai:20}. These combined figures of merit make PhCs uniquely suited for scalable chiral quantum optics, where both efficient photon extraction and deterministic spin-photon interfaces are required.  
\\
Beyond efficient interfaces, chiral coupling enables deterministic few-photon nonlinearities. Pairs of chirally coupled emitters can act as near-ideal photon sorters, separating single- and two-photon components with fidelities above 0.999, enabling deterministic non-linear sign gates and Bell-state analysis \cite{PRL128_213603}. Hybrid nanocavity-waveguide platforms further enhance functionality, combining strong Purcell enhancement ($F_{P} \approx 11$) with directional contrast ($\sim 88\%$) and wide-range electrical tunability, offering practical routes to electrically reconfigurable spin-photon interfaces \cite{Martin:25}. Together, these advances establish a clear foundation for scalable chiral networks, but achieving deterministic control over multiple emitters remains a central challenge that we address here.  

\begin{figure*}[ht]
\centering
\includegraphics[width=1\linewidth]{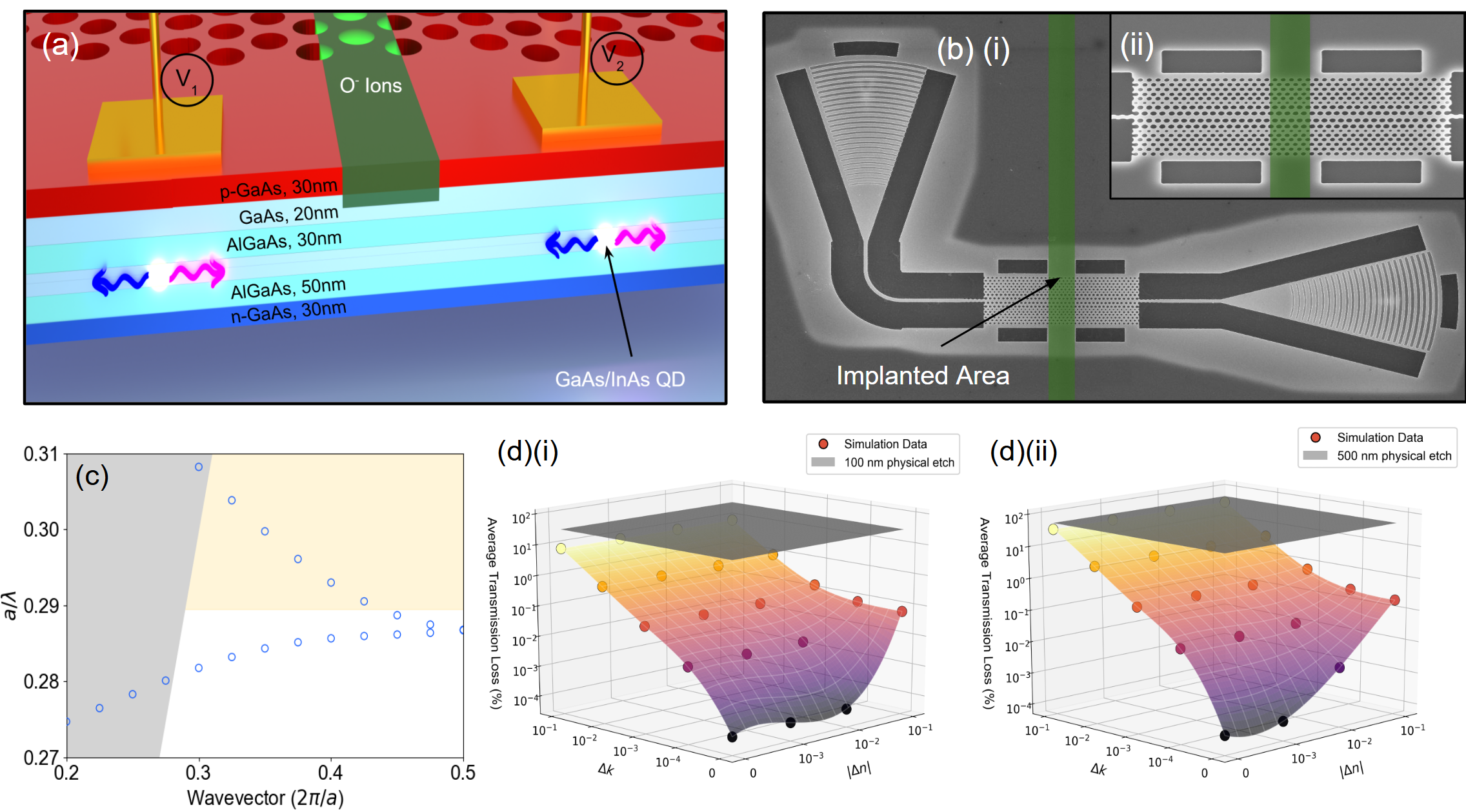}
\caption{(a) Schematic cross-section of \textit{p-i-n} membrane structure used for ion implantation. (b)(i) SEM image of the photonic crystal waveguide. (b)(ii) Top down SEM of GPWG highlighting the implanted region (500nm in width) seen as pseudo-colour green strip across the center of the waveguide.(c) Photonic band structure of the glide-plane waveguide; the shaded yellow region corresponds to the simulated band in (d). (d)(i, ii) Simulated transmission loss as a function of refractive-index ($\Delta n$) and extinction-coefficient ($\Delta k$) perturbations for 100nm and 500 nm implantation depths, respectively.}
\label{fig1}
\end{figure*}
Electrical tuning of QD transition energies via the quantum confined Stark effect (QCSE) offers a powerful solution to build programmable circuits for chiral quantum optics. It provides fast, reversible, and localized control of QD emission with compact integration. Furthermore it is the only approach that enables large scale synchronization of remote emitters, real-time stabilization of interference, and potential for dynamic wiring of graph states in chiral networks \cite{Buterakos2022Modular}. However, existing demonstrations achieve independent tuning by physically etching through the p-doped layers of a \textit{p--i--n} membrane, effectively splitting the device into two diodes\cite{PhysRevLett.131.033606,hallett2025controllingcoherencewaveguidecoupledquantum}. When introduced into a photonic crystal, such etching perturbs the local group index and increases both backscattering and out-of-plane losses\cite{mazoyer2009disorder}.  
\\
In this work, we overcome these challenges by using ion implantation, to provide independent electrical control of multiple emitters in GPWG's without compromising the advantages of the waveguide. Ion implantation provides lateral electrical isolation without physically etching the photonic crystal lattice, thereby preserving its optical properties. We extend on earlier demonstrations in photonic crystal nanocavities\cite{Bonato} and waveguide systems that introduced significant optical perturbations\cite{mi13020291}. Using this approach, we demonstrate control of two chirally coupled QDs while maintaining the high-chirality of the GPWG. Combining Stark tuning with Zeeman splitting, we selectively bring independent spin states ($\sigma^\pm$) from two QDs into resonance. By exploring photon correlations from different spin-state pairs, we highlight the potential of ion-implanted waveguides as a platform for stable, electrically reconfigurable chiral spin networks.  

\section{Methods}

The sample used in this work consists of a 170~nm GaAs membrane grown by molecular beam epitaxy (MBE) on top of a 1.1~\textmu m thick $\text{Al}_{0.6}\text{Ga}_{0.4}\text{As}$ sacrificial layer, supported by a GaAs substrate. The membrane is configured as a \textit{p-i-n} diode heterostructure, with a single layer of InAs quantum dots embedded at its center. The top 30~nm p-layer of the membrane is doped with carbon at a concentration of $2 \times 10^{19}~\text{cm}^{-3}$, while the bottom 30~nm is n-doped with silicon at a concentration of $2 \times 10^{18}~\text{cm}^{-3}$. In addition, there are two $\text{Al}_{0.3}\text{Ga}_{0.7}\text{As}$ barriers on either side of the QD layer. The wider band gap of AlGaAs suppresses carrier tunneling and leakage currents, allowing single QD Stark tuning of a few meV to 25 meV in QDs emitting at 900 nm using large AlGaAs barriers \cite{bennett2010giant}.

\begin{figure*}[ht]
\centering
\includegraphics[width=0.9\linewidth]{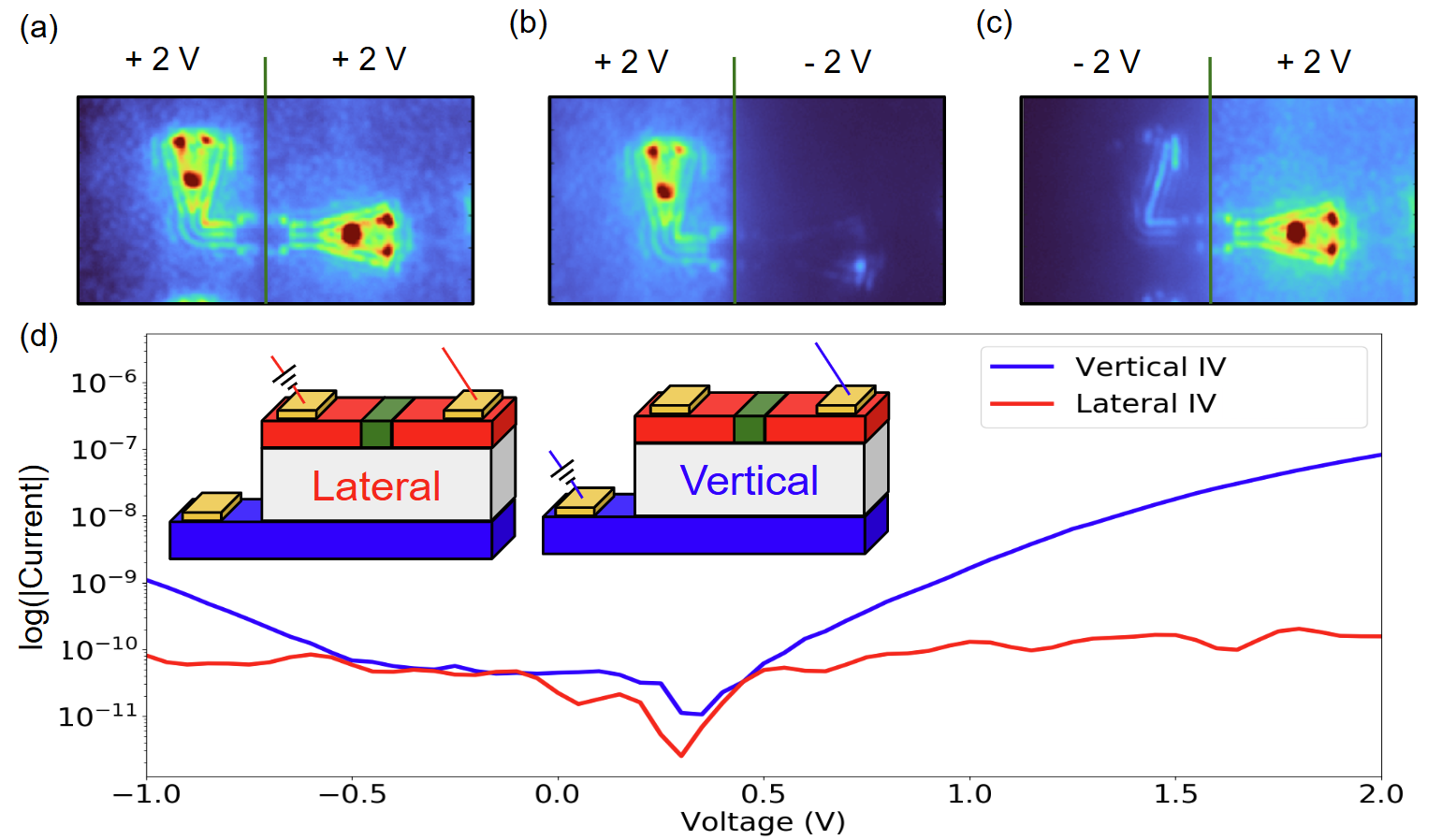}
\caption{(a-c) Confocal photoluminescence (PL) images showing emission from two electrically isolated regions of a diode under different applied biases. (a) Both left and right sections are forward biased at +2 V, resulting in emission across the full device. (b,c) Independent tuning demonstrated by locally forward biasing one section (+2 V) while reverse biasing the other (--$2$ V), leading to selective quenching of emission in the reverse-biased region. (d) Measured IV characteristics, comparing vertical IV across membrane (blue) and lateral IV across the implanted  barrier (red),showing strong suppression of lateral current.}
\label{fig:2}
\end{figure*}

Prior to fabrication of any photonic structures a SiO$_2$ hard mask is deposited and patterned using electron beam lithography (EBL), the exposed areas of the sample are then bombarded with O$^{-}$ ions at a dose of $2 \times 10^{12}~\text{cm}^{2}$ accelerated to 40~keV forming 500nm tracks across diode sections (The choice of width is justified in Supplementary.S1). The implantation is performed in a 7/22 orientation (tilt and azimuthal angle for in-plane rotation) to avoid ion channeling caused by alignment with the major crystal axis. The PhC is then patterned using a SiO$_2$ hard mask with EBL. The patterns are transferred into the GaAs membrane using inductively coupled plasma reactive ion etching (ICP-RIE). Electrical contacts to the p- and n-doped layers are made using Ti/Au metallization. This configuration allows for independent electrical biasing of the laterally separated regions of the diode by placing contact pads which can be wire bonded after full device fabrication and connected to a ceramic chip carrier. Following this, the wafer is submerged in hydrofluoric acid (HF) to selectively remove the sacrificial layer of $\text{Al}_{0.6}\text{Ga}_{0.4}\text{As}$, releasing the membrane and forming suspended photonic structures. To avoid mechanical damage during drying, a critical point drying process is employed.
\\
We apply this isolation technique to a glide-plane waveguide (SEM image shown in Fig.\ref{fig1}(b))\cite{Siampour2023}, chosen for its broken mirror symmetry, which enables highly chiral light-matter interactions. In these structures, spin-momentum locking allows the directional coupling of circularly polarized dipole transitions from quantum emitters into guided modes with near-unity efficiency\cite{coles2016chirality}. This chiral coupling is particularly pronounced in the vicinity of the photonic band edge, where the waveguide supports slow-light modes with high group index ($n_g \gg 1$). %In this regime, the group velocity is significantly reduced, leading to an enhanced local density of optical states and strong Purcell enhancement.
The slow-light region is therefore highly sensitive to perturbations in the refractive index profile: even minor dielectric discontinuities introduced by fabrication imperfections or material damage can induce backscattering, mode disruption, and substantial radiative losses. As such, the glide-plane waveguide serves as an ideal testbed for our electrical isolation method.
\\
Ion implantation achieves electrical isolation by modifying the dopant profile in the p-layer rather than physically altering the structure. This process introduces charge traps that form additional defect states with energies lying below the near-valence states of the p-doped GaAs. This immobilizes the free carriers, pinning the local Fermi level \cite{yin2022physical}. This pinning effect prevents modulation of the local electric field under applied bias, effectively decoupling the electric field across neighboring regions of the diode. As a result, the implanted section behaves as a high resistance barrier, screening the applied bias and preventing carrier transport. 
\\
To quantify the changes in optical properties resulting from ion implantation, we perform finite-difference time-domain (FDTD) simulations of unimplanted and implanted GPWGs [tidy 3d reference]. Based on SRIM simulations discussed further in supplementary S2, we confirm that the process operates well below the amorphization threshold, meaning we form defect states in the valence band without generating a continuous amorphous GaAsO$_x$ network that would otherwise lead to large variations in the complex refractive index $\tilde{n}$. Consequently, we estimate the change in $\tilde{n}$ from the corresponding reduction in charge mobility. This yields an approximate refractive-index change of $\Delta n \sim -10^{-3}$ and an extinction coefficient $\Delta k \sim 10^{-5}$.
\\
Fig.\ref{fig1}(c) presents the calculated photonic band structure of the GPWG, with the shaded yellow region indicating the guided mode of interest used in FDTD simulations. The transmission characteristics of this mode under various perturbation parameters are summarized in Fig.\ref{fig1}(d)(i) and \ref{fig1}(d)(ii). These plots show the simulated average transmission loss as a function of refractive-index change ($\Delta n$) and extinction coefficient ($\Delta k$) for implantation widths of 100~nm and $500~\text{nm}$, respectively.
\\
For small implantation widths of 100~nm, simulations reveal that the optical mode remains largely unaffected in the range we estimate to be operating in with perturbations of $|\Delta n| \sim 10^{-3}$ resulting in $<0.01\%$ of additional loss. In contrast, the black plane in Fig.\ref{fig1}(d)(i) represents the expected transmission loss of an equivalent 100~nm physical etch, exceeding 30\%. This comparison highlights the strong detrimental impact of physical etching on mode confinement and scattering. Moreover, at greater implantation widths of 500 nm Fig.\ref{fig1}(d)(ii), the guided mode continues to exhibit minimal sensitivity to the implanted region, showing only a marginal increase in transmission loss compared with the 100 nm case of $\sim0.04\%$. The corresponding black surface again represents a large optical loss associated with physical etching of the same width, which exceeds 50\%.

\begin{figure}[ht]
\centering
\includegraphics[width=\linewidth]{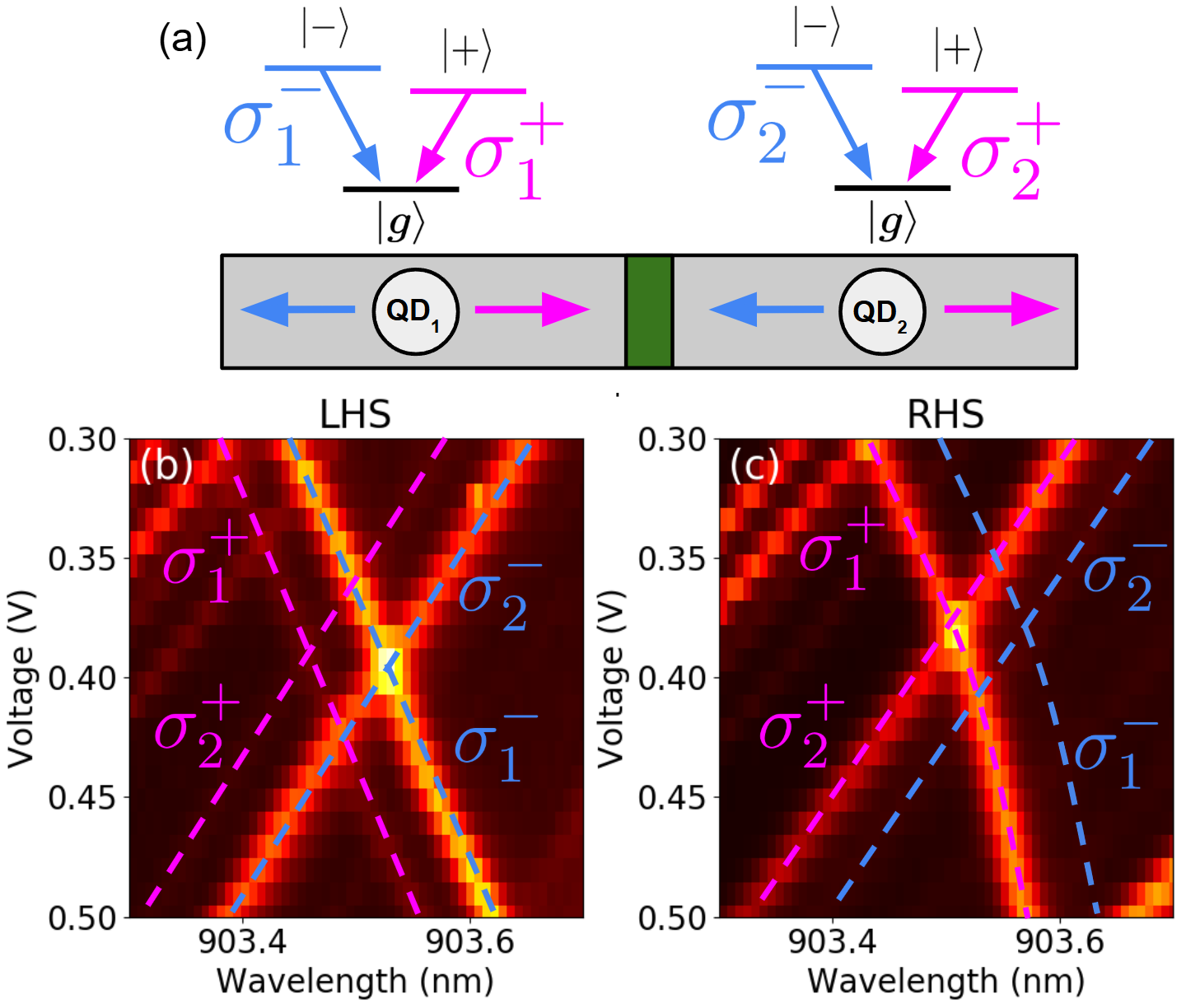}
\caption{(a) Schematic of the chiral PhC waveguide showing two spatially separated quantum dots (QD$_1$ and QD$_2$) embedded on either side of an implanted isolation barrier. The $\sigma^{+}$ and $\sigma^{-}$ spin transitions couple into opposite directions due to spin–momentum locking. (b,c)  PL maps acquired from the right propagating mode (RHS) and left propagating mode (LHS), showing Stark tuning of Zeeman split QD transitions under applied voltage. Colored dashed lines track the $\sigma^\pm$ transitions of each dot.}
\label{fig3}
\end{figure}

\section{Results}

In order to assess the effectiveness of the implanted section, we perform bias-dependent PL imaging and measure the IV characteristics of the device. Fig.~\ref{fig:2}(a–c) shows confocal PL images of an implanted device under different electrical bias configurations. When one side of the diode is held under reverse bias ($-2$V), QD emission is strongly quenched due to enhanced carrier tunnelling, resulting in the localized dark regions seen in Fig.~\ref{fig:2}(b–c). The ability to quench from either side of the device separately confirms effective lateral electrical isolation.  
\\
Fig.~\ref{fig:2}(d) illustrates the two measurement geometries used: (i) vertical IV, where current flows through the membrane, and (ii) lateral IV, where current flow is measured across the implanted region. The corresponding IV characteristics are shown in Fig.~\ref{fig:2}(e). The vertical IV exhibits typical \textit{p--i--n} diode behaviour, with an exponential increase in current above a turn-on voltage of $1.5\,\text{V}$. In contrast, the lateral IV across the implanted section shows a strongly suppressed current, remaining in the nA range even at forward biases where the vertical IV already carries $\mu$A currents. 
\\
At reverse biases beyond $-3\,\text{V}$, we observe breakdown behaviour in the lateral IV. From the slope of the steepest linear segment of the lateral IV trace in Fig.~\ref{fig:2}(e), and using the known device geometry, we estimate a lower bound on the effective resistivity of the implanted region to be approximately $1.4 \times 10^{5}\,\Omega\cdot\text{m}$. This corresponds to an increase in resistivity of around eight orders of magnitude compared to the unimplanted membrane, confirming that ion implantation provides robust electrical isolation without the need for physical etching. These findings are discussed in Supplementary.S3 with more detailed calculations.  
\\
Following this, to demonstrate how ion implantation allows for control of separate QD detuning while maintaining the optical properties of a PhC waveguide, we perform PL measurements on two chirally coupled QDs in a glide-plane waveguide. We apply a 0.5~T magnetic field in the Faraday geometry, which lifts the spin degeneracy of excitonic states of QDs in the waveguide, giving rise to circularly polarized transitions ($\sigma^{+}$ and $\sigma^{-}$) that directionally couple into opposite propagation directions of the waveguide mode as illustrated in Fig.\ref{fig3}(a).
\\
The resulting PL maps, shown in Fig.\ref{fig3}(b,c), are collected from both ends of the PhC via shallow-etch grating couplers that terminate the device. The bias on the diode closest to the right-hand-side (RHS) outcoupler is held constant at 1.43~V, while the bias applied to the left-hand-side (LHS) diode is varied. Each Zeeman-split transition exhibits clear and continuous Stark tuning, confirming minimal spectral diffusion or charge noise. Notably, no abrupt spectral jumps or drift are observed over the full tuning range, attesting to the high stability of the implanted waveguide environment. This level of spectral control is essential for the realization of indistinguishable photon sources and coherent spin–photon interfaces in scalable quantum photonic circuits. Notably, we observe signs of cross-talk between the two QDs, as evident in the PL maps. This behavior is attributed to the physical proximity of the QDs to the diode interface ($\sim$ 3 and 5 $\mu m$ for LHS and RHS of implantation strip respectively). Despite this, the data confirm that ion implantation provides sufficient electrical isolation to enable differential tuning of closely spaced emitters within the same photonic device.
\\
Importantly, we do not observe any substantial degradation of directionality as the emission propagate across the implanted region separating the two diode sections. The emission collected from the LHS and RHS remains strongly spin-selective, with only a minor asymmetry in relative intensity between the two directions, comparable to that observed in unimplanted reference devices. QD1 has an averaged asymmery of $1.82 \pm 0.75\%$ and $12.76\% \pm 1.6\%$ This minor deviation could be attributed to weak residual backscattering at the implantation interface or subtle fabrication imperfections, rather than any intrinsic modification of the underlying photonic mode structure, and is consistent with similar asymmetries reported even in nominally symmetric photonic crystal systems where such effects are not expected\cite{Martin:25,Martin2023Topological,JalaliMehrabad}.
\\
This contrasts strongly with the case of physically etched glide-plane waveguides, where FDTD simulations predict a pronounced reduction in chirality from $\sim98\%$ for an unetched device to $\sim38\%$ for etch widths of $500~\text{nm}$. The absence of such degradation in the implanted structures demonstrates that low-dose oxygen implantation modifies the refractive index only perturbatively, without disrupting the guided mode symmetry or the deterministic spin–photon transfer essential for chiral quantum interfacing.

\begin{figure*}[ht]
\centering
\includegraphics[width=\linewidth]{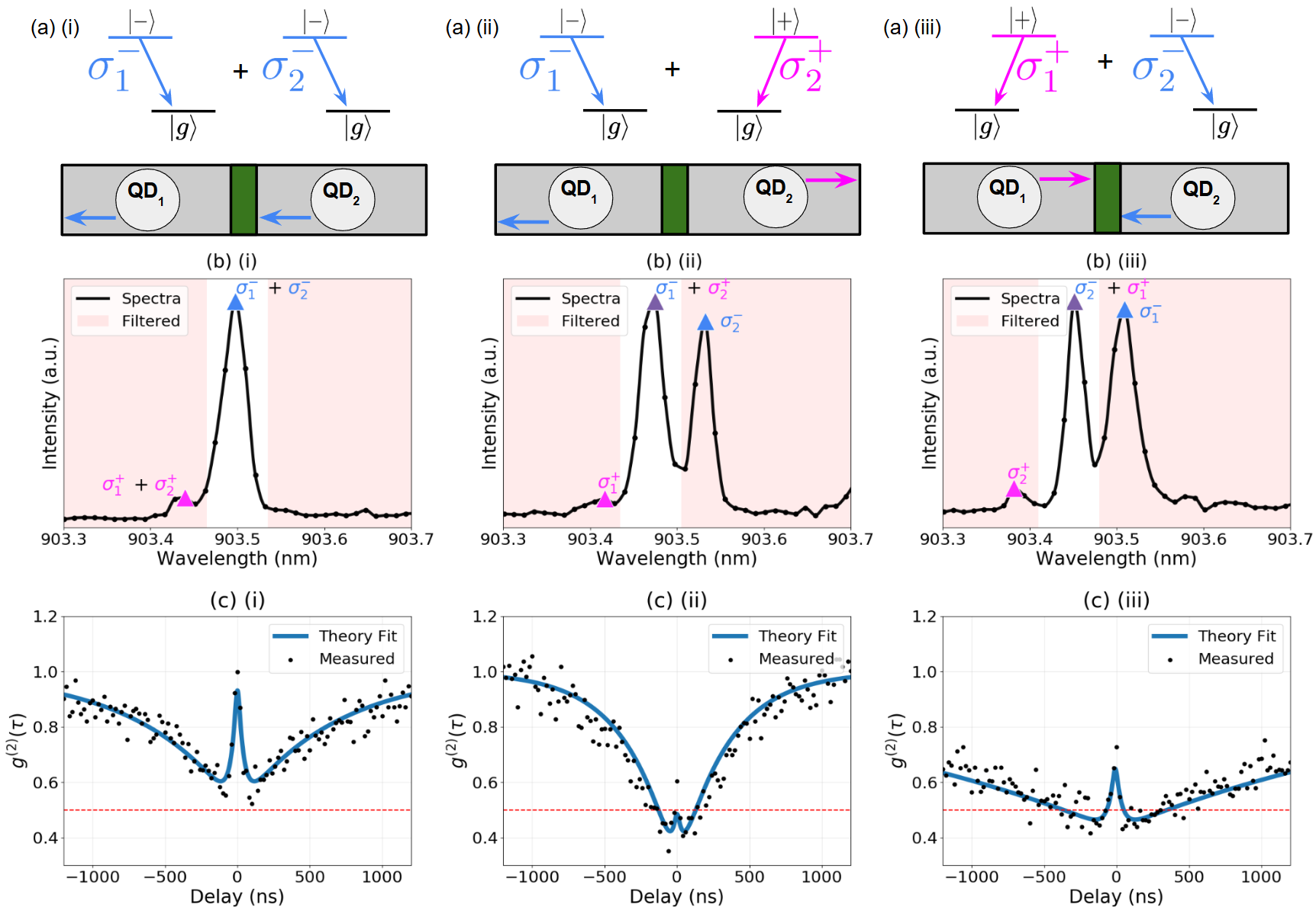}
\caption{(a) Spin configurations for three two-emitter correlation measurements where light is collected from the left hand side of the waveguide shown in Fig.~\ref{fig3}(b)
(i) the $\sigma^{-}$ of QD1 and QD2 are measured, (ii) the $\sigma^{-}$  state of QD1 and $\sigma^{+}$  of QD2 are measured, (iii) the $\sigma^{+}$  state of QD1 and $\sigma^{-}$  of QD2 are measured (b) Photoluminescence spectra demonstrating the tuning of various spin pairs into resonance. (i) $ \sigma^{-}_{1}$ + $ \sigma^{-}_{2}$, (ii) $ \sigma^{-}_{1}$ + $ \sigma^{+}_{2}$, (iii) $ \sigma^{-}_{2}$ + $ \sigma^{+}_{1}$, with red-shaded regions indicating the spectral windows used for filtering prior to correlation measurements. (c) Second-order photon correlation functions $g^{(2)}(\tau)$ for the spin state combinations in (a), showing theoretical fits (solid lines) and experimental data (black dots). The quantum limit, $g^{(2)}(\tau)=0.5$, is indicated by the red dashed line.}
\label{fig4}
\end{figure*}

Finally, we probe spin-selective emission and evaluate quantum interference between the two emitters using second-order photon correlation (Hanbury Brown–Twiss, HBT) measurements. This technique is particularly powerful for assessing the long-term spectral stability of quantum dots, since indistinguishable photons will interfere and produce clear bunching or antibunching signatures only if their emission energies remain stable within the photon coherence time \cite{Santori2002}. Any spectral diffusion or Stark jitter would wash out these correlations, leading to a degraded visibility. Thus, the observation of robust spin-dependent photon correlations directly demonstrates that ion implantation maintains spectral stability over extended measurement times.
\\
Beyond stability, HBT measurements provide a stringent test of indistinguishability, an essential requirement for applications in quantum photonic networks. In chiral quantum systems, where spin–momentum locking routes photons of opposite helicity into distinct propagation channels, interference between indistinguishable emitters enables deterministic entanglement generation, entanglement swapping, and quantum repeater functionality. Demonstrating stable and selective spin–photon correlations therefore highlights the suitability of our non-destructive tuning method for scalable architectures that rely on multi-emitter interference in chiral photonic circuits.
\\
To assess these points, we perform second-order photon correlation measurements in the same optical pumping and collection scheme shown in Fig.\ref{fig3}(b). For each measurement, a different pair of transitions of the two QDs are tuned to resonance, and a 70pm band-pass filter is used to isolate emission from the resonant lines. The three spin-pair configurations studied are shown in Fig.\ref{fig4}(a,b): (i) $\sigma_-$+$\sigma_-$, (ii) $\sigma_-$+$\sigma_+$, and (iii) $\sigma_+$+$\sigma_-$ for QD$_1$ and QD$_2$, respectively. The corresponding autocorrelation functions $g^{(2)}(\tau)$ are plotted in Fig.\ref{fig4}(c). The three measurements have been fitted to a theoretical model, using a consistent set of parameters across the measurements (discussed further in Supplementary.S4).
\\
In configuration Fig.\ref{fig4}(c)(i) where the two spin states are coupled to the same propagating direction, strong photon bunching is observed at zero time delay, consistent with two-photon interference between indistinguishable photons emitted into the same mode with equal contribution. In contrast, configurations Fig.\ref{fig4}(c)(ii) and Fig.\ref{fig4}(c)(iii) where the two spin states measured are coupled predominantly to opposing propagating directions show reduced or absent bunching, reflecting the directional separation of opposite helicities in the chiral waveguide. As photons of opposing spin are routed into different optical modes, they do not interfere at the beam splitter as frequently, reducing the coincidence rate at $\tau = 0$. The observed spin-dependent correlation signatures reveal both the spin-state interference and the underlying chiral symmetry of the QDs in the waveguide. These results confirm selective spin-photon coupling, robust spectral stability, and coherent interference between remote emitters, all essential components for scalable chiral quantum networks.
\\
\section{Discussion and Outlook}
We demonstrate a scalable method for achieving electrical tuning of spatially separated QDs embedded in a photonic crystal waveguide. The implantation of targeted ions into the p-doped region of a \textit{p--i--n} diode creates a high resistivity barrier that enables Stark tuning while preserving the optical properties of the surrounding nanophotonic structure. Unlike physical etching, which introduces detrimental scattering and degrades photonic mode integrity, our approach maintains the high-quality light–matter interaction essential for scalable chiral quantum networks.
\\
PL imaging and I–V measurements confirm robust electrical isolation, while correlation experiments demonstrate long-term spectral stability and spin-resolved addressability of the quantum dots. In particular, spin-resolved second-order correlation measurements, $g^{(2)}(\tau)$, reveal chirality-dependent photon bunching with negligible Stark-induced spectral jitter. This stability confirms that our isolation scheme supports extended experimental operation, a crucial requirement for protocols involving long measurement times, high-visibility interference, and quantum network synchronization \cite{Senellart2017,Warburton2013}.
\\
These results establish spin–photon interfacing between indistinguishable emitters mediated by the chiral waveguide—a critical milestone toward functional multi-emitter quantum networks. More broadly, the demonstrated stability and addressability make this approach attractive for applications such as deterministic entanglement generation \cite{Somaschi2016}, quantum repeaters, and chiral quantum optical interfaces.
\\
Our work highlights ion implantation as a powerful tool for electrically isolating and tuning multiple emitters on the same chip, fully compatible with photonic integrated circuit architectures. By optimizing the implantation dosage and implementing an electrical contacting scheme with a dedicated n-doped contact layer per diode section, full and independent tuning of individual emitters can be achieved more reliably. The development of such non-destructive isolation methods thus represents a key step toward scalable multi-qubit operations and integrated quantum photonic circuits.
\\
\\
\textbf{Funding} : This work was supported by EPSRC Grant No.EP/N031776/1 and EP/V026496/1 and the Quantum
Communications Hub EP/T001011/1.

\newpage
\bibliography{Bib_Main}

@article{mahmoodian2016quantum,
  title = {Quantum Networks with Chiral-Light--Matter Interaction in Waveguides},
  author = {Mahmoodian, Sahand and Lodahl, Peter and S\o{}rensen, Anders S.},
  journal = {Phys. Rev. Lett.},
  volume = {117},
  issue = {24},
  pages = {240501},
  numpages = {6},
  year = {2016},
  month = {Dec},
  publisher = {American Physical Society},
  doi = {10.1103/PhysRevLett.117.240501},
  url = {https://link.aps.org/doi/10.1103/PhysRevLett.117.240501}
}

@Article{Niu2023,
author={Niu, Daoheng
and Zhang, Yuxuan
and Shabani, Alireza
and Shapourian, Hassan},
title={All-photonic one-way quantum repeaters with measurement-based error correction},
journal={npj Quantum Information},
year={2023},
month={Oct},
day={21},
volume={9},
number={1},
pages={106},
issn={2056-6387},
doi={10.1038/s41534-023-00775-9},
url={https://doi.org/10.1038/s41534-023-00775-9}
}

@Article{Madsen2022Borealis,
author={Madsen, Lars S.
and Laudenbach, Fabian
and Askarani, Mohsen Falamarzi.
and Rortais, Fabien
and Vincent, Trevor
and Bulmer, Jacob F. F.
and Miatto, Filippo M.
and Neuhaus, Leonhard
and Helt, Lukas G.
and Collins, Matthew J.
and Lita, Adriana E.
and Gerrits, Thomas
and Nam, Sae Woo
and Vaidya, Varun D.
and Menotti, Matteo
and Dhand, Ish
and Vernon, Zachary
and Quesada, Nicol{\'a}s
and Lavoie, Jonathan},
title={Quantum computational advantage with a programmable photonic processor},
journal={Nature},
year={2022},
month={Jun},
day={01},
volume={606},
number={7912},
pages={75-81},
issn={1476-4687},
doi={10.1038/s41586-022-04725-x},
url={https://doi.org/10.1038/s41586-022-04725-x}
}

@Article{Yang2025Programmable,
author={Yang, Yang
and Chapman, Robert J.
and Youssry, Akram
and Haylock, Ben
and Lenzini, Francesco
and Lobino, Mirko
and Peruzzo, Alberto},
title={Programmable quantum circuits in a large-scale photonic waveguide array},
journal={npj Quantum Information},
year={2025},
month={Feb},
day={03},
volume={11},
number={1},
pages={19},
issn={2056-6387},
doi={10.1038/s41534-024-00934-6},
url={https://doi.org/10.1038/s41534-024-00934-6}
}

@article{Slussarenko,
    author = {Slussarenko, Sergei and Pryde, Geoff J.},
    title = {Photonic quantum information processing: A concise review},
    journal = {Applied Physics Reviews},
    volume = {6},
    number = {4},
    pages = {041303},
    year = {2019},
    month = {10},
    issn = {1931-9401},
    doi = {10.1063/1.5115814},
    url = {https://doi.org/10.1063/1.5115814},
}

@Article{Tomm2021,
author={Tomm, Natasha
and Javadi, Alisa
and Antoniadis, Nadia Olympia
and Najer, Daniel
and L{\"o}bl, Matthias Christian
and Korsch, Alexander Rolf
and Schott, R{\"u}diger
and Valentin, Sascha Ren{\'e}
and Wieck, Andreas Dirk
and Ludwig, Arne
and Warburton, Richard John},
title={A bright and fast source of coherent single photons},
journal={Nature Nanotechnology},
year={2021},
month={Apr},
day={01},
volume={16},
number={4},
pages={399-403},
issn={1748-3395},
doi={10.1038/s41565-020-00831-x},
url={https://doi.org/10.1038/s41565-020-00831-x}
}

@Article{Somaschi2016,
author={Somaschi, N.
and Giesz, V.
and De Santis, L.
and Loredo, J. C.
and Almeida, M. P.
and Hornecker, G.
and Portalupi, S. L.
and Grange, T.
and Ant{\'o}n, C.
and Demory, J.
and G{\'o}mez, C.
and Sagnes, I.
and Lanzillotti-Kimura, N. D.
and Lema{\'i}tre, A.
and Auffeves, A.
and White, A. G.
and Lanco, L.
and Senellart, P.},
title={Near-optimal single-photon sources in the solid state},
journal={Nature Photonics},
year={2016},
month={May},
day={01},
volume={10},
number={5},
pages={340-345},
issn={1749-4893},
doi={10.1038/nphoton.2016.23},
url={https://doi.org/10.1038/nphoton.2016.23}
}

@Article{Meng2024,
author={Meng, Yijian
and Chan, Ming Lai
and Nielsen, Rasmus B.
and Appel, Martin H.
and Liu, Zhe
and Wang, Ying
and Bart, Nikolai
and Wieck, Andreas D.
and Ludwig, Arne
and Midolo, Leonardo
and Tiranov, Alexey
and S{\o}rensen, Anders S.
and Lodahl, Peter},
title={Deterministic photon source of genuine three-qubit entanglement},
journal={Nature Communications},
year={2024},
month={Sep},
day={05},
volume={15},
number={1},
pages={7774},
issn={2041-1723},
doi={10.1038/s41467-024-52086-y},
url={https://doi.org/10.1038/s41467-024-52086-y}
}

@article{Arcari2014NearUnity,
  title = {Near-Unity Coupling Efficiency of a Quantum Emitter to a Photonic Crystal Waveguide},
  author = {Arcari, M. and S\"ollner, I. and Javadi, A. and Lindskov Hansen, S. and Mahmoodian, S. and Liu, J. and Thyrrestrup, H. and Lee, E. H. and Song, J. D. and Stobbe, S. and Lodahl, P.},
  journal = {Phys. Rev. Lett.},
  volume = {113},
  issue = {9},
  pages = {093603},
  numpages = {5},
  year = {2014},
  month = {Aug},
  publisher = {American Physical Society},
  doi = {10.1103/PhysRevLett.113.093603},
  url = {https://link.aps.org/doi/10.1103/PhysRevLett.113.093603}
}

@article{Lodahl2015rmp,
  title = {Interfacing single photons and single quantum dots with photonic nanostructures},
  author = {Lodahl, Peter and Mahmoodian, Sahand and Stobbe, S\o{}ren},
  journal = {Rev. Mod. Phys.},
  volume = {87},
  issue = {2},
  pages = {347-400},
  numpages = {54},
  year = {2015},
  month = {May},
  publisher = {American Physical Society},
  doi = {10.1103/RevModPhys.87.347},
  url = {https://link.aps.org/doi/10.1103/RevModPhys.87.347}
}

@Article{Sollner2015Deterministic,
author={S{\"o}llner, Immo
and Mahmoodian, Sahand
and Hansen, Sofie Lindskov
and Midolo, Leonardo
and Javadi, Alisa
and Kir{\v{s}}ansk{\.{e}}, Gabija
and Pregnolato, Tommaso
and El-Ella, Haitham
and Lee, Eun Hye
and Song, Jin Dong
and Stobbe, S{\o}ren
and Lodahl, Peter},
title={Deterministic photon--emitter coupling in chiral photonic circuits},
journal={Nature Nanotechnology},
year={2015},
month={Sep},
day={01},
volume={10},
number={9},
pages={775-778},
issn={1748-3395},
doi={10.1038/nnano.2015.159},
url={https://doi.org/10.1038/nnano.2015.159}
}

@article{QuanLoncar2011,
author = {Qimin Quan and Marko Loncar},
journal = {Opt. Express},
number = {19},
pages = {18529--18542},
publisher = {Optica Publishing Group},
title = {Deterministic design of wavelength scale, ultra-high Q photonic crystal nanobeam cavities},
volume = {19},
month = {Sep},
year = {2011},
url = {https://opg.optica.org/oe/abstract.cfm?URI=oe-19-19-18529},
doi = {10.1364/OE.19.018529},
}

@misc{germanis2025waveguideexcitationspinpumping,
      title={Waveguide Excitation and Spin Pumping of Chirally Coupled Quantum Dots}, 
      author={Savvas Germanis and Xuchao Chen and René Dost and Dominic J. Hallett and Edmund Clarke and Pallavi K. Patil and Maurice S. Skolnick and Luke R. Wilson and Hamidreza Siampour and A. Mark Fox},
      year={2025},
      eprint={2502.00218},
      archivePrefix={arXiv},
      primaryClass={cond-mat.mes-hall},
      url={https://arxiv.org/abs/2502.00218}, 
}

@Inbook{SKgrowth,
author={Rastelli, Armando
and Kiravittaya, Suwit
and Schmidt, Oliver G.},
editor={Michler, Peter},
title={Growth and control of optically active quantum dots},
bookTitle={Single Semiconductor Quantum Dots},
year={2009},
publisher={Springer Berlin Heidelberg},
address={Berlin, Heidelberg},
isbn={978-3-540-87446-1},
doi={10.1007/978-3-540-87446-1_2},
url={https://doi.org/10.1007/978-3-540-87446-1_2}
}

@article{Zhai:20,
author = {Liang Zhai and Matthias C. Löbl and Giang N. Nguyen and Julian Ritzmann and Alisa Javadi and Clemens Spinnler and Andreas D. Wieck and Arne Ludwig and Richard J. Warburton},
journal = {Nature Communications},
title = {Low-noise GaAs quantum dots for quantum photonics},
year = {2020},
volume = {11},
number = {1},
pages = {4745},
month = {Sep},
url = {https://doi.org/10.1038/s41467-020-18625-z},
doi = {10.1038/s41467-020-18625-z},
}

@article{Martin2023Topological,
  title = {Topological and conventional nanophotonic waveguides for directional integrated quantum optics},
  author = {Martin, N. J. and Jalali Mehrabad, M. and Chen, X. and Dost, R. and Nussbaum, E. and Hallett, D. and Hallacy, L. and Foster, A. and Clarke, E. and Patil, P. K. and Hughes, S. and Hafezi, M. and Fox, A. M. and Skolnick, M. S. and Wilson, L. R.},
  journal = {Phys. Rev. Res.},
  volume = {6},
  issue = {2},
  pages = {L022065},
  numpages = {8},
  year = {2024},
  month = {Jun},
  publisher = {American Physical Society},
  doi = {10.1103/PhysRevResearch.6.L022065},
  url = {https://link.aps.org/doi/10.1103/PhysRevResearch.6.L022065}
}

@article{Hauff2022Chiral,
  title = {Chiral quantum optics in broken-symmetry and topological photonic crystal waveguides},
  author = {Hauff, Nils Valentin and Le Jeannic, Hanna and Lodahl, Peter and Hughes, Stephen and Rotenberg, Nir},
  journal = {Phys. Rev. Res.},
  volume = {4},
  issue = {2},
  pages = {023082},
  numpages = {16},
  year = {2022},
  month = {Apr},
  publisher = {American Physical Society},
  doi = {10.1103/PhysRevResearch.4.023082},
  url = {https://link.aps.org/doi/10.1103/PhysRevResearch.4.023082}
}

@article{PRL128_213603,
  title = {Deterministic Photon Sorting in Waveguide QED Systems},
  author = {Yang, Fan and Lund, Mads M. and Pohl, Thomas and Lodahl, Peter and M\o{}lmer, Klaus},
  journal = {Phys. Rev. Lett.},
  volume = {128},
  issue = {21},
  pages = {213603},
  numpages = {6},
  year = {2022},
  month = {May},
  publisher = {American Physical Society},
  doi = {10.1103/PhysRevLett.128.213603},
  url = {https://link.aps.org/doi/10.1103/PhysRevLett.128.213603}
}

@article{Martin:25,
author = {Nicholas J. Martin and Dominic Hallett and Mateusz Duda and Luke Hallacy and Elena Callus and Luke Brunswick and Ren\'{e} Dost and Edmund Clarke and Pallavi K. Patil and Pieter Kok and Maurice S. Skolnick and Luke R. Wilson},
journal = {Optica},
number = {7},
pages = {1100--1108},
publisher = {Optica Publishing Group},
title = {Purcell-enhanced, directional light\&\#x2013;matter interaction in a waveguide-coupled nanocavity},
volume = {12},
month = {Jul},
year = {2025},
url = {https://opg.optica.org/optica/abstract.cfm?URI=optica-12-7-1100},
doi = {10.1364/OPTICA.561630},
}

@article{JalaliMehrabad,
author = {Mahmoud Jalali Mehrabad and Andrew P. Foster and Ren\'{e} Dost and Edmund Clarke and Pallavi K. Patil and A. Mark Fox and Maurice S. Skolnick and Luke R. Wilson},
journal = {Optica},
number = {12},
pages = {1690--1696},
publisher = {Optica Publishing Group},
title = {Chiral topological photonics with an embedded quantum emitter},
volume = {7},
month = {Dec},
year = {2020},
url = {https://opg.optica.org/optica/abstract.cfm?URI=optica-7-12-1690},
doi = {10.1364/OPTICA.393035},
}

@article{Buterakos2022Modular,
  doi = {10.22331/q-2023-03-02-935},
  url = {https://doi.org/10.22331/q-2023-03-02-935},
  title = {Modular architectures to deterministically generate graph states},
  author = {Shapourian, Hassan and Shabani, Alireza},
  journal = {{Quantum}},
  issn = {2521-327X},
  publisher = {{Verein zur F{\"{o}}rderung des Open Access Publizierens in den Quantenwissenschaften}},
  volume = {7},
  pages = {935},
  month = mar,
  year = {2023}
}

@misc{hallett2025controllingcoherencewaveguidecoupledquantum,
      title={Controlling coherence between waveguide-coupled quantum dots}, 
      author={D. Hallett and J. Wiercinski and L. Hallacy and S. Sheldon and R. Dost and N. Martin and A. Fenzl and I. Farrer and A. Verma and M. Cygorek and E. M. Gauger and M. S. Skolnick and L. R. Wilson},
      year={2025},
      eprint={2410.17890},
      archivePrefix={arXiv},
      primaryClass={quant-ph},
      url={https://arxiv.org/abs/2410.17890}, 
}

@article{mazoyer2009disorder,
  title = {Disorder-induced incoherent scattering losses in photonic crystal waveguides: Bloch mode reshaping, multiple scattering, and breakdown of the Beer-Lambert law},
  author = {Patterson, M. and Hughes, S. and Schulz, S. and Beggs, D. M. and White, T. P. and O'Faolain, L. and Krauss, T. F.},
  journal = {Phys. Rev. B},
  volume = {80},
  issue = {19},
  pages = {195305},
  numpages = {6},
  year = {2009},
  month = {Nov},
  publisher = {American Physical Society},
  doi = {10.1103/PhysRevB.80.195305},
  url = {https://link.aps.org/doi/10.1103/PhysRevB.80.195305}
}

@article{Bonato,
    author = {Thon, S. M. and Kim, H. and Bonato, C. and Gudat, J. and Hagemeier, J. and Petroff, P. M. and Bouwmeester, D.},
    title = {Independent electrical tuning of separated quantum dots in coupled photonic crystal cavities},
    journal = {Applied Physics Letters},
    volume = {99},
    number = {16},
    pages = {161102},
    year = {2011},
    month = {10},
    issn = {0003-6951},
    doi = {10.1063/1.3651491},
    url = {https://doi.org/10.1063/1.3651491},
    eprint = {https://pubs.aip.org/aip/apl/article-pdf/doi/10.1063/1.3651491/14458479/161102\_1\_online.pdf},
}

@Article{mi13020291,
AUTHOR = {Yu, Xingshi and Chen, Xia and Milosevic, Milan M. and Shen, Weihong and Topley, Rob and Chen, Bigeng and Yan, Xingzhao and Cao, Wei and Thomson, David J. and Saito, Shinichi and Peacock, Anna C. and Muskens, Otto L. and Reed, Graham T.},
TITLE = {Ge Ion Implanted Photonic Devices and Annealing for Emerging Applications},
JOURNAL = {Micromachines},
VOLUME = {13},
YEAR = {2022},
NUMBER = {2},
ARTICLE-NUMBER = {291},
URL = {https://www.mdpi.com/2072-666X/13/2/291},
PubMedID = {35208415},
ISSN = {2072-666X},
DOI = {10.3390/mi13020291}
}

@article{bennett2010giant,
  title={Giant Stark effect in the emission of single semiconductor quantum dots},
  author={Bennett, Anthony J and Patel, Raj B and Skiba-Szymanska, Joanna and Nicoll, Christine A and Farrer, Ian and Ritchie, David A and Shields, Andrew J},
  journal={Applied Physics Letters},
  volume={97},
  number={3},
  year={2010},
  publisher={AIP Publishing}
}

@Article{Siampour2023,
author={Siampour, Hamidreza
and O'Rourke, Christopher
and Brash, Alistair J.
and Makhonin, Maxim N.
and Dost, Ren{\'e}
and Hallett, Dominic J.
and Clarke, Edmund
and Patil, Pallavi K.
and Skolnick, Maurice S.
and Fox, A. Mark},
title={Observation of large spontaneous emission rate enhancement of quantum dots in a broken-symmetry slow-light waveguide},
journal={npj Quantum Information},
year={2023},
month={Feb},
day={22},
volume={9},
number={1},
pages={15},
issn={2056-6387},
doi={10.1038/s41534-023-00686-9},
url={https://doi.org/10.1038/s41534-023-00686-9}
}

@Article{coles2016chirality,
author={Coles, R. J.
and Price, D. M.
and Dixon, J. E.
and Royall, B.
and Clarke, E.
and Kok, P.
and Skolnick, M. S.
and Fox, A. M.
and Makhonin, M. N.},
title={Chirality of nanophotonic waveguide with embedded quantum emitter for unidirectional spin transfer},
journal={Nature Communications},
year={2016},
month={Mar},
day={31},
volume={7},
number={1},
pages={11183},
issn={2041-1723},
doi={10.1038/ncomms11183},
url={https://doi.org/10.1038/ncomms11183}
}

@article{yin2022physical,
title = {Physical mechanism of field modulation effects in ion implanted edge termination of vertical GaN Schottky barrier diodes},
journal = {Fundamental Research},
volume = {2},
number = {4},
pages = {629-634},
year = {2022},
issn = {2667-3258},
doi = {https://doi.org/10.1016/j.fmre.2021.11.027},
url = {https://www.sciencedirect.com/science/article/pii/S2667325821002855},
author = {Ruiyuan Yin and Chiachia Li and Bin Zhang and Jinyan Wang and Yunyi Fu and Cheng P. Wen and Yilong Hao and Bo Shen and Maojun Wang}
}

@Article{Santori2002,
author={Santori, Charles
and Fattal, David
and Vu{\v{c}}kovi{\'{c}}, Jelena
and Solomon, Glenn S.
and Yamamoto, Yoshihisa},
title={Indistinguishable photons from a single-photon device},
journal={Nature},
year={2002},
month={Oct},
day={01},
volume={419},
number={6907},
pages={594-597},
issn={1476-4687},
doi={10.1038/nature01086},
url={https://doi.org/10.1038/nature01086}
}

@Article{Senellart2017,
author={Senellart, Pascale
and Solomon, Glenn
and White, Andrew},
title={High-performance semiconductor quantum-dot single-photon sources},
journal={Nature Nanotechnology},
year={2017},
month={Nov},
day={01},
volume={12},
number={11},
pages={1026-1039},
issn={1748-3395},
doi={10.1038/nnano.2017.218},
url={https://doi.org/10.1038/nnano.2017.218}
}

@Article{Warburton2013,
author={Warburton, Richard J.},
title={Single spins in self-assembled quantum dots},
journal={Nature Materials},
year={2013},
month={Jun},
day={01},
volume={12},
number={6},
pages={483-493},
issn={1476-4660},
doi={10.1038/nmat3585},
url={https://doi.org/10.1038/nmat3585}
}

@article{PhysRevLett.131.033606,
  title = {Independent Electrical Control of Two Quantum Dots Coupled through a Photonic-Crystal Waveguide},
  author = {Chu, Xiao-Liu and Papon, Camille and Bart, Nikolai and Wieck, Andreas D. and Ludwig, Arne and Midolo, Leonardo and Rotenberg, Nir and Lodahl, Peter},
  journal = {Phys. Rev. Lett.},
  volume = {131},
  issue = {3},
  pages = {033606},
  numpages = {6},
  year = {2023},
  month = {Jul},
  publisher = {American Physical Society},
  doi = {10.1103/PhysRevLett.131.033606},
  url = {https://link.aps.org/doi/10.1103/PhysRevLett.131.033606}
}
\end{document}